\documentclass[prb,twocolumn,showpacs]{revtex4-2}

\usepackage[utf8]{inputenc} 
\usepackage{graphicx}
\usepackage{ifthen}
\usepackage{ifpdf}
\usepackage{color}
\definecolor{red}{rgb}{1.,0.,0.}
\definecolor{blue}{rgb}{0.,0.,1.}

\usepackage{latexsym}
\usepackage{amsmath}
\usepackage{bm}
\usepackage{bbm} 
\usepackage{soul}
\newcommand{\half}{{\textstyle\frac{1}{2}}}

\newcommand{\im}{\mbox{Im}}

\newcommand{\eexp}{\mbox{e}^}

\newcommand{\AB}{{\bf A}}

\newcommand{\rr}{{\bf r}}

\newcommand{\qq}{{\bf q}}
\newcommand{\kk}{{\bf k}}

\newcommand{\nn}{{\bf n}}

\newcommand{\beq}[1]{\begin{eqnarray}\ifthenelse{#1=-1}{\nonumber}
		{\ifthenelse{#1=0}{}{\label{e#1}}}}
	\newcommand{\eeq}{\end{eqnarray}}
\newcommand{\be}{\begin{equation}}
\newcommand{\ee}{\end{equation}}

\newcommand{\bea}{\begin{eqnarray}}
\newcommand{\eea}{\end{eqnarray}}

\newcommand{\hide}[1]{}

\newcommand{\sign}{\mathop{\rm sign}}

\usepackage{lineno}
\usepackage{lipsum}

\begin{document}


\title{Randomly twisted bilayer graphene: Cascade transitions}
\author{Baruch Horovitz$^{1}$ and Pierre Le Doussal$^{2}$}
\affiliation{$^1$Department of Physics, Ben Gurion University of the Negev, Beer Sheva 84105, Israel\\
	{$^2$} Laboratoire de Physique de l'Ecole Normale Sup\'erieure, CNRS, ENS and PSL Universit\'e, Sorbonne Universit\'e, Universit\'e Paris Cit\'e, 24 rue Lhomond, 75005 Paris, France.}
\begin{abstract}
	Twisted bilayer graphene (TBG) is known to have disorder in its twist angle. We show that in terms of a Dirac equation with a random gauge potential $\AB(\rr)$ this disorder becomes huge when the average twist angle is near the magic angle where the Dirac velocity vanishes. The density of states (DOS) then diverges at the Dirac point as $\rho(E)\sim E^{(2/z)-1}$ with $z\gg 1$ and we deduce that electrons occupy energies very near $E=0$. We solve a strong disorder problem by deriving a sum rule on the disorder averaged eigenfunctions from which we deduce that each added electron contributes equal intraband Coulomb interaction energy. The various bands in TBG are related by either $\AB(\rr)\rightarrow \AB(-\rr)$ or $\AB(\rr)\rightarrow -\AB(\rr)$ which affects the interband interaction energy. We find, within Hartree-Fock, jumps in the chemical potential at each integer filling, accounting for the cascade transitions. 
\end{abstract}
\maketitle

\section{Introduction} 

Twisted bilayer graphene (TBG) is extensively studied as a unique platform for strongly correlated electrons. It has been shown that near the corners of the Moire Brillouin zone the electrons obey a 2D Dirac equation with  a velocity that vanishes \cite{BM} at a magic twist angle of $\theta_m\approx 1.1^\circ$. By approaching $\theta_m$ the interaction vs kinetic energies is easily tuned.

The remarkable cascade transitions \cite{ilani1,yacobi1,herrero,saito} show a sequence of peaks in the inverse compressibility $d\mu/d\nu$ when the filling factor of electrons in a Moir\'e unit cell is near an integer $|\nu|=0,1,2,3$,
 corresponding to several flavors, each having a Dirac point. We note that the shape of the cascade transitions depends markedly on the size of the probed system, a pattern that recurs in symmerically twisted trilayer graphene (TTG) \cite{yacobi2,liu}. The peaks are broad and shifted from integer $\nu$ when probing $\approx 0.5 \mu$m \cite{ilani1} or $\approx 0.1 \mu$ in TTG \cite{yacobi2}, while they are sharper and more accurately at integer $\nu$  when probing larger areas by either directly averaging over $\approx 2 \mu$m \cite{yacobi1} or by a global measurements over a few $\mu$m \cite{herrero,saito} and up to $\approx 10 \mu$m in TTG \cite{liu}. 

We propose that this size dependence is related to the disorder in the twist angle. The latter was studied in TBG and shown to be present on scales $\gtrsim 0.5 \mu$m \cite{zeldov}. There has been a considerable theoretical progress in understanding TBG and its cascade transitions \cite{BM,ilani1,herrero,rai}, however, none of these theories includes twist angle disorder. We claim that disorder average is essential in any long scale measurement such that as those global experiments or in transport data.

In the present work we  solve a strong disorder problem. We first show that the twist angle disorder generates a random gauge potential $\AB(\rr)$ within a 2D Dirac equation. It can be taken as transverse \cite{footnotetransverse}, i.e. $A_x=\partial_y V$, $A_y=-\partial_x V$,
with potential $V$ and effective magnetic field  ${\sf B}=\nabla^2 V$. The disorder
average, $\overline{{\sf B}(\qq) {\sf B}(\qq')} = \pi \sigma q^2 (2 \pi)^2 \delta(\qq+\qq')$
in Fourier space, defines the dimensionless disorder parameter $\sigma$, which,
as we show, becomes huge near the magic angle.
It is known that gauge disorder produces an anomalous density of states (DOS) \cite{ludwig} $\rho(E)\sim E^{(2/z)-1}$, which undergoes a freezing transition, and becomes divergent at large $\sigma$, with $z\sim\sqrt{\sigma}$ \cite{bh}.
We infer then that the electrons occupy states very near $E=0$ for each of the flavours.

Furthermore, we prove a sum rule for a disorder average of the eigenfunctions of the Dirac equation, from which we deduce that each added electron carries equal average Coulomb interaction energy. Using known symmetries and explicitly known $E=0$ solutions we deduce the average interaction also between the various flavours. Finally we evaluate the Hartree-Fock interaction energy and show that the chemical potential $\mu$ jumps at each integer $\nu$, accounting for the main feature of the cascade transitions. The appendices provide details on symmetries and on the reduced Hamiltonian for both TBG and TTG, on density overlaps of the explicitly known E=0 states and on disorder averages of $\rho^2$ that lead to the sum rules.
\\

\section{Twist angle disorder}

 We parametrize the deviation of the twist angle from its mean value as $\delta\theta(\rr)$, and its disorder average by
\beq{01}
\overline{\delta\theta(\rr+\delta\rr)\delta\theta(\rr)}=\gamma(\frac{a}{L_m})^2\eexp{-|\delta\rr|/\ell_0}
\eeq
where $a,\,L_m$ are the lattice constants of a single layer and of the bilayer Moir\'e lattice, respectively, 
with $L_m \approx 13 nm$, and $\ell_0$ is a correlation length. Ref. \onlinecite{zeldov} provides detailed analysis of the $\delta\theta(\rr)$ distribution, showing two samples with standard deviations of 0.025$^{\circ}$ and 0.022$^{\circ}$. Since $\frac{a}{L_m}=2\sin\half\theta\approx \theta_m$ we infer $\gamma\approx 5\cdot 10^{-4}$. Assuming that $\ell_0$ is dominated by the maximal ${\bm\nabla}\theta$ these samples imply $\ell_0=0.13-0.21\mu$m, alternatively taking their "typical gradient" of 0.05$^{\circ}/\mu$m implies $\ell_0\approx 0.5\mu$m. The onset of Landau fans at $B\approx $1 T, corresponding to a cyclotron resonance of 30 nm, indicates a shorter $\ell_0$.

Consider next the resulting strain.  A single graphene layer whose unit cell is tilted by an angle $\delta\theta(\rr)$ has a shear strain \cite{NelsonHalperin} 
\beq{02}
\half  (\partial_x u_y-\partial_y u_x) = \delta\theta
\eeq
This relation involves the transverse component of the displacement pattern $(u_x,\,u_y)$, 
whose Fourier components are $u_\alpha(\qq)= 2  i \epsilon_{\alpha\beta}q_\beta 
\delta\theta(\qq)/q^2$.
The strain implies that the nearest neighbor C-C spacing $a'=a/\sqrt{3}$ as well as the corresponding transfer integral are modulated, hence the Dirac equation acquires a gauge vector potential \cite{paco1,paco2,koshino,mele} 
\beq{03}
&&\tilde A_x(\qq)=i\frac{\beta}{2a'}(q_xu_x-q_yu_y)
=-\frac{{2} \beta}{a'}\frac{q_xq_y}{q^2} \delta\theta\nonumber\\&&
\tilde A_y(\qq)={ -  i\frac{\beta}{2a'}(q_yu_x+q_xu_y)}
=\frac{\beta}{a'}\frac{q_y^2-q_x^2}{q^2} \delta\theta
\eeq
where $\beta\approx 3$ (Ref. \cite{paco1}),
leading to a fictitious magnetic field ${\sf B}(\qq) =  i q_x \frac{3 q_y^2-q_x^2}{q^2 } \frac{\beta}{a'} \delta \theta(\qq)$. Upon angular averaging for the scale $\ell_0\gg L_m$ we can identify the disorder parameter as
\beq{04}
&&\pi\tilde\sigma\equiv\int d^2r \sum_{j=x,y}\overline{ \tilde A_j(\rr) \tilde A_j(\rr')}\nonumber\\&&
={ \frac{1}{2}} (\frac{\beta}{a'})^2\int d^2r \overline{ \delta\theta(\rr)\delta\theta(0)} \approx {27 \pi} \gamma(\frac{\ell_0}{L_m})^2 
\eeq
Each layer has actually $\half\delta\theta$ with $\gamma=5\cdot 10^{-4}/4$, hence for the range $\ell_0=200-500$nm we estimate  $\tilde \sigma=0.8-5$ 
per layer. 
We note that the strain due to lattice relaxation \cite{paco2,mele} is also described by a gauge potential which is, however, periodic on the Moir\'e lattice, while in our case $\AB(\rr)$ varies on the scale $\ell_0\gg L_m$.

Consider next a one layer Hamiltonian $v_F{\bm\sigma}\cdot\kk$ with velocity $v_F$, Pauli matrices ${\bm\sigma}$ and momentum $\kk$ measured from a chosen Moir\'e Brillouin zone corner (a Dirac point). The perturbative approach \cite{BM,paco2,mele} (appendix A) couples this Dirac point by wavevectors $\qq_i$ to the three neighboring corners that correspond to states on the other layer. This interlayer tunnelling is $u_0$ on AA regimes and $u_1$ on AB regimes (AA or AB refer to the relative arrangement of the two layers). 
The interlayer coupling contributes an opposite sign $\sim -v_F{\bm\sigma}\cdot\kk$, hence the effective velocity $v$ vanishes at $\theta_m$. Including $\tilde\AB(\rr)$ in this derivation is valid since $\ell_0\gg L_m$, i.e. the $\qq_i$ apply on a scale where $\tilde\AB(\rr)$ is locally constant. Since $\tilde\AB(\rr)\sim\delta\theta(\rr)$ has opposite signs on the two layers the near cancellation of $v_F$ does not apply to $\tilde\AB(\rr)$ leading to
an effective Hamiltonian 
\beq{05}
&&{\cal H}_{eff}=v{\bm \sigma}\cdot [-i{\bm \nabla}-\AB(\rr)]\nonumber\\
&&v=v_F\frac{1-3\bar u_1^2}{1+3(\bar u_0^2+\bar u_1^2)},\,\, \AB(\rr)=\frac{1+3\bar u_1^2}{1-3\bar u_1^2}\tilde\AB(\rr)
\eeq
where $\bar u_i=u_i\frac{3a}{4\pi v_F\theta},\,i=1,2$.
The velocity $v$ coincides with the original \cite{BM} result when $u_0=u_1$. It indicates the presence of a magic angle, and is in fact very close to the exact (numerical) result \cite{BM,paco2,mele} for $\AB=0$. The same form of ${\cal H}_{eff}$ is found also for symmetric TTG (appendix A).
We note that in deriving ${\cal H}_{eff}$ factors $\eexp{\pm i\theta\sigma_z/4}$ that can lead to non-abelian disorder \cite{fosterW} are neglected here and elsewhere \cite{paco2,mele,bernevig} since $\theta$ is small (appendix A).

It is remarkable that at the magic angle $3\bar u_1^2\rightarrow 1$ the coefficient of $\tilde\AB(\rr)$ remains finite, thus the disorder effect becomes stronger and would induce a divergent DOS \cite{ludwig} 
and a freezing transition \cite{bh} before reaching the magic angle. 
An expansion in $\theta$ yields $v(\theta+\delta\theta)=v(\theta)+ v_F\frac{\delta\theta}{\theta}$; with the given standard deviation we consider $\delta\theta= 0.025^{\circ}$ 
as the closest approach to $\theta_m$, the velocity is then reduced to $v\approx v_F\,0.025/1.1 \approx v_F/40$. At this situation the disorder parameter $ \pi\sigma=\int d^2r \sum_{j=x,y}\overline{ A_j(\rr) A_j(\rr')}$ is enhanced to $\sigma=40^2\tilde\sigma$, i.e. given the previous estimate $\sigma$ can reach huge values of $10^3-10^4$.

We proceed to identify the various flavors and their symmetry relations (also in appendix A) related to the symmetries of the ordered system \cite{bernevig,sheffer}. The operator $\eta_x$ for exchanging the two layers results in the same Hamiltonian Eq. \eqref{e05} except $\AB(\rr)$ changes sign, hence $\eta_x=\{\AB(\rr)\rightarrow -\AB(\rr)\}$. This symmetry relates the two Moir\'e valleys, a flavor denoted as $v_m=\pm$. Consider next states built from distinct graphene valleys $v=\pm$ via complex conjugation and space reflection $R$, hence ${\bm\sigma}\rightarrow{\bm\sigma}^*,\,\AB(\rr)\rightarrow \AB(-\rr)$. Adding finally the electron spin $s=\pm$ we end up with three flavors $v_m,\,v,\,s$ and therefore eight bands, consistent with the $\AB(\rr)=0$ case \cite{sachdev}. A significant operation is the particle hole (p-h) symmetry, $C=\eta_x\sigma_x \cal K$ where $\cal K$ is complex conjugation, since 
$C{\cal H}_{eff}C=-{\cal H}_{eff}$, similar to the $\AB(\rr)=0$ case \cite{bernevig,sheffer}.  We also note a time reversal symmetry operator $\sigma_y\eta_y{\cal K}$ and a chiral symmetry $\sigma_z\eta_z$ as in similar cases \cite{fosterX}. Thus our system of 8 Dirac points corresponds to class BDI.

The $E=0$ explicit eigenfunctions are known \cite{ludwig}, hence the symmetries determine these also for the other flavors. 
In table \ref{table1} we summarize the symmetry relations, the Hamiltonian of each flavor and the exact $E=0$ eigenstates in terms of $V(\rr)$,  where we recall that $A_x(\rr)= \partial_y V(\rr),\,A_y(\rr)=-\partial_xV(\rr)$, 
the spin variable $s$ is implicit, unaffecting the Hamiltonians. Each Hamiltonian is $2\times 2$ in its subspace, i.e. ${\bm\sigma}$ act on the AA and AB states of the Moir\'e unit cell, thus the whole 3-flavor system is a $16\times 16$ matrix. 
Table \ref{table1} shows two eigenstates for each of the flavors, corresponding to $E\rightarrow \pm 0$, related by the p-h operator $C$. The third column in the table is chosen to represent $E\rightarrow -0$ states as generated from each other by the symmetries in the first column, while the fourth column represents $E\rightarrow +0$ states (the 3rd and 4th columns may be interchanged as a whole, unaffecting the following results).

 For simplified reading one may proceed directly to Eqs. \eqref{e12} and \eqref{e17} for just the $E=0$ states as can be straighforwardly derived, and then proceed to the Hartree-Fock section. The purpose of the next sections on DOS and sum rules is to prove that Eq. \eqref{e17} is valid for all eigenstates.

\begin{widetext}
	
	\begin{table}[t]
		\caption{ Flavors and eigenstates}
		\def\arraystretch{1.7}
			\begin{tabular}{|c|c|c|c|}
				Symmetry & Flavor Hamiltonians & Eigenstates  & Eigenstates  \\ relation & ${\cal H}_{v_m,v}$  & $E\rightarrow -0$ & $E\rightarrow +0$ \\ \hline
				1 & ${\cal H}_{+,+}=v{\bm \sigma}\cdot[-i{\bm\nabla}-\AB(\rr)]$  &
				$\begin{pmatrix} \eexp{V(\rr)} \\ 0\end{pmatrix}$ & $\begin{pmatrix} 0\\ \eexp{-V(\rr)}\end{pmatrix}$  \\ 
				$\eta_x$ & ${\cal H}_{-,+}=v{\bm\sigma} \cdot[-i{\bm\nabla}+\AB(\rr)]$ &
				$\begin{pmatrix} \eexp{-V(\rr)} \\ 0\end{pmatrix}$ & $\begin{pmatrix} 0\\ \eexp{V(\rr)}\end{pmatrix}$  \\
				$\sigma_x R$  & ${\cal H}_{+,-}=v{\bm\sigma}^* \cdot[i{\bm\nabla}-\AB(-\rr)]$ &
				$\begin{pmatrix} 0\\ \eexp{V(-\rr)} \end{pmatrix}$ & $\begin{pmatrix}  \eexp{-V(-\rr)} \\ 0\end{pmatrix}$  \\
				$\sigma_x R\eta_x$ & ${\cal H}_{-,-}=v{\bm\sigma}^* \cdot[i{\bm\nabla}+\AB(-\rr)]$ &
				$\begin{pmatrix} 0 \\ \eexp{-V(-\rr)} \end{pmatrix}$ & $ \begin{pmatrix}  \eexp{V(-\rr)} \\ 0\end{pmatrix}$\\
				\hline 
			\end{tabular}
		\label{table1}
	\end{table}
	
\end{widetext}

\section{DOS}

 The local DOS is found by projecting the fermion action into the subspace of energy $E+i\epsilon$ ($E,\,\epsilon$ real) and then bosonizing \cite{ludwig,bh}. The disorder is averaged with the replica method, generating boson fields 
$\phi_a(\rr)$ with $a=1,...,m$, eventually $m\rightarrow 0$. The non-interacting bosonic action is
\beq{06}
&&{\cal S}_0=\int d^2r\frac{1}{8\pi}\sum_{a,b}(K^{-1}\delta_{a,b}+\sigma){\bm\nabla}\phi_a\cdot{\bm\nabla}\phi_b
\eeq
where the $K$ parameter is a conventional generalization \cite{giam}, though up to now $K=1$. The system is solved with $E=0$ where the bare interaction $\frac{\epsilon}{2\pi \alpha v}\sum_a\int d^2r\eexp{i \phi_a(\rr)}+h.c.$ involves a cutoff $\alpha$ of the fermion-boson transformation  \cite{giam}, $\alpha\approx L_m$.  Since the disorder sets in at a longer scale $\ell_0$ we first integrate out wavevectors in the range $1/\ell_0<q<1/\alpha$ so that 
$\langle\eexp{i\phi_a(\rr)}\rangle_0=(\frac{\alpha}{\ell_0})^K\eexp{i\phi_a(\rr)}$ with $\phi_a(\rr)$ now varies on scales $\gtrsim \ell_0$. Thus all states correspond now to wavevectors $q\lesssim 1/\ell_0$ or densities $\lesssim 1/\ell_0^2$. The bare interaction becomes 
$\sum_a\int\frac{d^2r}{\ell_0^2}\tilde\epsilon\eexp{i\phi_a(\rr)}+h.c.$ where
$\tilde\epsilon=\frac{\epsilon\alpha}{2\pi v}(\frac{\alpha}{\ell_0})^{K-2}$.
We note that deviations from a linear dispersion at $q>1/\ell_0$ could change the definition of $\tilde\epsilon$. In the following we use $\tilde\epsilon$ as the input for further disorder renormalization on scales $q<1/\ell_0$.

To treat properly the strong disorder case and obtain the correct scaling dimension of  $\tilde\epsilon\cos\phi_a(\rr)$ we include from the start all higher order terms as generated by renormalization group \cite{scheidl, carpentier,carpentier2, doussal,bh2}. The interaction is then  $\sum_{\{\nn\}}Y[\nn]\eexp{i\nn\cdot{\bm\phi}}$ where $\{\nn\}$ are a set of vectors of length $m$ and entries $0,\pm 1$, ${\bm \phi}=\{\phi_1(\rr),...,\phi_m(\rr)\}$ and $Y[\nn]=\Pi_a\tilde\epsilon^{\,n_a^2}$; the bare interaction has $\sum_a n_a^2=1$. The bosonic action is then
\beq{07}
{\cal S}_B={\cal S}_0-\int\frac{d^2r}{\ell_0^2}
\sum_{\{\nn\}}Y[\nn]\eexp{i\nn\cdot{\bm\phi}}
\eeq

The DOS, either in terms of the eigenfunctions of the Dirac Hamiltonian $\varphi_n(\rr)$ and their eigenvalues $E_n$ or in terms of the bosonic variables \cite{bh}, is 
\beq{08}
\overline{\rho_\epsilon (E,\rr)}=&&-\frac{1}{\pi}\im \overline{ \sum_n \frac{1}{E-E_n+i\epsilon}|\varphi_n(\rr)|^2} \nonumber\\&&
=\frac{1}{\pi\ell_0^2}\frac{\partial}{\partial\epsilon}\langle\sum_{\{\nn\}}Y[\nn]\eexp{i\nn\cdot{\bm\phi}}\rangle
\eeq
where $\langle...\rangle$ averages $\phi_a(\rr)$ with $S_B$, the notation $|\varphi_n(\rr)|^2\equiv \varphi_n^\dagger(\rr)\cdot\varphi_n(\rr)$ means a spinor product. 
We evaluate 
$\rho_\epsilon(\rr)=\rho_\epsilon(E=0,\rr)$ with a variational method and summarize here the results \cite{bh}. The interaction is replaced by   $\sum_{a,b} \half[\sigma_c(K,\tilde\epsilon)\delta_{ab}+\sigma_0(K,\tilde\epsilon)]\phi_a(\rr)\phi_b(\rr)$ and the variational condition determines  $\sigma_0(K,\tilde\epsilon)\lesssim\sigma_c(K,\tilde\epsilon)\approx \ell_0^{-2}\tilde\epsilon^{2/z}$ up to a numerical prefactor. The exponent is
$z=K(\sqrt{8\sigma}-1)$ for strong disorder $\sigma>2/K^2$, while $z=2-K+\sigma K^2$ for weak disorder. 
 We note that in the classical limit ($K=0$, any $\sigma$), where $\cos\phi_a(\rr)$ is expanded near its minimum, $\sigma_c(K=0,\tilde\epsilon)=\frac{\epsilon}{2\pi v\alpha}$ is correctly reproduced. Furthermore, for $\sigma=0$ $\sigma_c(K,\tilde\epsilon)$ is $\ell_0$ independent, as it should.
 
The average DOS \eqref{e08} is then 
\beq{09} \label{09} 
\overline{\rho_\epsilon(\rr)}\approx\frac{1}{2\pi^2 v\alpha}(\frac{\alpha}{\ell_0})^K\tilde\epsilon^{\,(2/z)-1}
\eeq
To find the conventional DOS $\rho(E,\rr)=\sum_n\delta(E-E_n)|\varphi_n(\rr)|^2$ we note the relation
\beq{10} 
&&\rho_\epsilon(\rr)=\frac{1}{\pi}\int dE \frac{\epsilon}{\epsilon^2+E^2}\rho(E,\rr)\nonumber\\&&
\Rightarrow \overline{\rho(E,\rr)}\approx\frac{\sin(\pi/z)}{2\pi^2 v \alpha}
(\frac{\alpha}{\ell_0})^K\tilde  E^{\,(2/z)-1}
\eeq
where $\tilde E=\frac{E\alpha}{2\pi v}(\frac{\alpha}{\ell_0})^{K-2} $, hence the DOS diverges at $E=0$ when $z>2$. The average density at a chemical potential $\mu$ is then
\beq{11}
\overline{\rho(\rr)}&&=\int_0^\mu \overline{\rho(E,\rr)} dE=\frac{\sin \pi/z}{\pi\ell_0^2}
\int_0^{\tilde\mu} \tilde E^{(2/z)-1}d\tilde E\nonumber\\&&\approx \frac{1}{2\ell_0^2}\tilde\mu^{2/z} 
\eeq
where in the last relation $z\gg 1$ and $\tilde \mu=\frac{\mu\alpha}{2\pi v}(\frac{\alpha}{\ell_0})^{K-2} $. Hence $\mu$ is flat for the admissible densities 
$\overline{\rho(\rr)}\lesssim 1/\ell_0^2$, reflecting the divergent DOS.
\hide{We note that the average density follows a crossover function $\overline{\rho(\rr)}=\frac{1}{2L^2} f[\tilde\mu (\frac{L}{\ell_0})^z]$ where $f(0)=1$ and  $f(x)\rightarrow x^{2/z}$ at $x\gg 1$, interpolating between single particle and high density cases.}
\\

\section{Sum rules}

 Our aim in this section is to evaluate density overlaps, i.e. disorder average $\overline{|\varphi_i(\rr)|^2 |\varphi_j(\rr)|^2}$ for the various 2-spinor eigenstates of the Dirac equation \eqref{e05}. We consider first the $E=0$ states listed in table \ref{table1}.
These are evaluated in appendix B (including their normalization factors) using replica methods, aiming to find their $L$ dependence. The results in terms of the scaler $V(\rr)$ are, for $\sigma>\sigma_c=2$
\beq{12}
&&(a)\qquad \overline{\int d^2r \eexp{4V(\rr)}/N_+^2}\approx \frac{1}{\ell_0^2}\nonumber\\&& (b) \qquad
\overline{\int d^2r/N_+ N_- }\approx (\frac{L}{\ell_0})^{ 2-8\sqrt{\sigma/2}}\frac{1}{\ell_0^2}\rightarrow 0
\nonumber\\&& (c) \qquad
\overline{\int d^2r\eexp{[2V(\rr)\pm 2V(-\rr)]}/N_+ N_\pm }\approx \frac{1}{L^2}
\eeq
where the normalization factors are $N_\pm=\int d^2r \eexp{\pm 2V(\rr)}$.
These results determine readily the 64 density overlaps between all the states listed in table \ref{table1}. We note that (\ref{e12}a) is actually the inverse participation ratio, that has been studied by various methods \cite{carpentier,chamon,fyodorov}, consistent {with} the present result. We note further that (\ref{e12}b) shows that the intensity overlap of each state with its p-h conjugate, as well as same energy overlap of states with $v_m=\pm$ and same $v,\,s$ are all vanishingly small.

We proceed to evaluate density overlaps of all $E\neq 0$ 
states by proving sum rules, derived by squaring both the fermionic and bosonic forms of \eqref{e08} and then averaging. Consider first the fermionic form of \eqref{e08}
\beq{13}
\rho_\epsilon^2(\rr)=&&\frac{1}{\pi^2}\sum_{i,j}\frac{\epsilon^2}{(E_i^2+\epsilon^2)(E_j^2+\epsilon^2)} |\varphi_i(\rr)|^2|\varphi_j(\rr)|^2\nonumber\\&&\approx
\frac{1}{\pi^2\epsilon^2}\sum_{i,j}^{|E_i|,|E_j|<\epsilon}
|\varphi_i(\rr)|^2|\varphi_j(\rr)|^2\nonumber\\
\eeq
so that $\epsilon$ can be interpreted as the chemical potential $\mu$. The disorder average $\overline{\rho_\epsilon^2}$ in its bosonic form is in fact a 2-layer problem \cite{bh} as detailed in appendix C. We find
\beq{14}
\overline{\rho_\epsilon^2}=(\frac{1}{\pi\ell_0^2 })^2\frac{\ell_0^2}{\epsilon^2}\sigma_c(2K,\tilde\epsilon^2)\approx \frac{2 \overline{\rho(\rr)}}{\pi^2\ell_0^2\epsilon^2}
\eeq
where the exponent $z=K(\sqrt{8\sigma}-1)\rightarrow 2z$ so that $\sigma_c(2K,\tilde\epsilon^2)\approx\ell_0^{-2}(\tilde\epsilon^2)^{2/2z}=
\ell_0^{-2}\tilde\mu^{\, 2/z}=2\overline{\rho(\rr)}$ using Eq. \eqref{e11}. 
 We note that $\overline{\rho_\epsilon^2}=\frac{1}{L^2\ell_0^2\epsilon^2} f[\tilde\epsilon(\frac{L}{\ell_0})^z]$ satisfies a scaling form with $f(0)=1,\, f(x)\sim x^{2/z}$ at $x\gg 1$, interpolating between the single particle (\ref{e12}a) and the finite density case \eqref{e14}. A similar scaling for class AIII was proposed \cite{fosterC}, though for individual terms of Eq. \eqref{e13}.

Equating the two forms (\ref{e13},\ref{e14}) yields a sum rule on the disorder average of all occupied states of the Dirac equation
\beq{15}
\hspace{-1cm} \sum_{i,j}^{|E_i|,|E_{j}|<\mu}
\int d^2r\overline{|\varphi_i[\AB(\rr),\rr)]|^2|\varphi_j[\AB(\rr),\rr]|^2}\approx \frac{N}{\pi\ell_0^2}
\eeq
where $\overline{\rho(\rr)} L^2=N$ is the total number of electrons in the partial filled band with $E\geq 0$; $\AB(\rr)$ is specified here to allow for other overlaps below. We note that with the single $E=0$ state the result with $N=1$  reproduces Eq. (\ref{e12}a), the latter was derived without using the fermion-boson transformation. 
The change $N\rightarrow N+1$ corresponds to an addition of one overlap in \eqref{e15} with either $E_i=\pm E_j=\mu$. The sign choice is prompted by the known $E=0$ overlap and the shared symmetry of either $E>0$ or $E<0$ states. Since at equal $E\rightarrow 0$ the $E_i=E_j$ overlap is dominant while $E_i=-E_j$ is vanishingly small (Eq. (\ref{e12}b)) we deduce that all the terms in \eqref{e15} are equal to $\delta_{i,j}/\pi\ell_0^2$.

Consider next the density overlaps between the states of ${\cal H}_{++}$ and ${\cal H}_{-+}$, generated by $\overline{\rho_\epsilon[\AB(\rr)]\rho_\epsilon[-\AB(\rr)]}$. This case also leads to a similar sum rule (interchanging $\phi_a^\pm$ in Eq. \eqref{e81})
\beq{16}
\hspace{-1cm} \sum_{i,j}^{|E_i|,|E_j|<\mu}
\int d^2r\overline{|\varphi_i [\AB(\rr),\rr]|^2|\varphi_j [-\AB(\rr),\rr]|^2}\approx \frac{N}{\pi\ell_0^2}
\eeq
In this case $E\rightarrow +0$ state of ${\cal H}_{++}$ in table \ref{table1} strongly overlaps with the $E\rightarrow -0$ state of ${\cal H}_{-+}$ while equal energy states have a vanishingly small overlap. Thus we deduce in this case that the change $N\rightarrow N+1$ generates an $E_i=-E_j$ overlap and that all terms in \eqref{e16} are equal to $\delta_{i,-j}/\pi\ell_0^2$, where $\delta_{i,-j}$ means $E_i=-E_j$.

Next consider the overlaps of ${\cal H}_{++}$  with ${\cal H}_{+-}$ states. The latter involve  $\rr\rightarrow -\rr$, hence the overlap is that of states localized at different positions and we expect  
$\int d^2r\overline{|\varphi_i[\AB(\rr),\rr]|^2|\varphi_j[\AB(-\rr),-\rr]|^2}\approx \delta_{i,j}/L^2$
consistent with Eq. (\ref{e12}c). In this case, however, we do not have an exact sum rule; yet this can be motivated by considering the overlap of densities $\eexp{2V(\rr)}$ and $\eexp{2V(\rr-\rr_0)}$ with $\rr_0\gg\ell_0$, representing states localized at different positions. This overlap is evaluated in Eq. \eqref{e70} and yields the same result as Eq. (\ref{e12}c). Same reasoning and result apply to overlaps of ${\cal H}_{++}$  with ${\cal H}_{--}$ states.

Defining interaction parameters by multiplying the various overlaps by $U\ell_0^2$ (to be used in the next section) we finally obtain
\beq{17}
&& U_1=U\ell_0^2\int d^2r\overline{|\varphi_i[\AB(\rr),\rr)]|^2|\varphi_j[\AB(\rr),\rr]|^2}\approx 
U\delta_{i,j}\nonumber\\&& U_1= U\ell_0^2
\int d^2r\overline{|\varphi_i[\AB(\rr),\rr)]|^2|\varphi_j[-\AB(\rr),\rr]|^2}\approx
U\delta_{i,-j}\nonumber\\&& U_2=U\ell_0^2
\int d^2r\overline{|\varphi_i[\AB(\rr),\rr)]|^2|\varphi_j[\AB(-\rr),-\rr]|^2}\approx
U\frac{\ell_0^2}{L^2}\delta_{i,j}\nonumber\\&&
\eeq
These results are fully consistent with the various overlaps in Eq. \eqref{e12}  of the $E=0$ states. For the latter states we can find the matrix elements also for small $\sigma$, see Eq. \eqref{e71}.
\vspace{2cm}

\begin{widetext}
	
	\begin{table*}[t]
		\caption{HF flavor filling and chemical potentials}
		\begin{center}
			\begin{tabular}{|c|c|c|}
				$\qquad\nu\qquad $ &  states  & $\qquad\mu\qquad$\\ 
				\hline
				-4 & $\qquad \uparrow ++,\,\downarrow -+\, E<0\qquad$ & 0 \\
				-3 & $\uparrow +-,\, \downarrow --\, E<0$ & $2U_2$\\
				-2 & $\qquad \downarrow ++,\,\uparrow -+\, E<0\qquad$ & $U_1+2U_2$ \\
				-1 & $\qquad \downarrow +-,\,\uparrow --\, E<0\qquad$ & $U_1+4U_2$ \\
				\hline
				0 & $\qquad \uparrow ++,\,\downarrow -+\, E>0\qquad$ & $2U_1+4U_2$ \\
				1 & $\qquad \uparrow +-,\,\downarrow --\, E>0\qquad$ & $2U_1 +6U_2$  \\
				2 & $\qquad \downarrow ++,\,\uparrow -+\, E>0\qquad$ & $3U_1+6U_2 $ \\
				3 & $\qquad \downarrow +-,\,\uparrow --\, E>0\qquad$ & $\qquad 3U_1+8U_2 \qquad$\\
				\hline 
			\end{tabular}
		\end{center}
		\label{table2}
	\end{table*}
	
\end{widetext}

\section{Hartree Fock}

 The Coulomb interaction $V_c(\rr)$ has a screening length \cite{ilani1} of $d/\epsilon_B\approx 10$nm, where d is the distance to the gate the $\epsilon_B$ is the dielectric constant of the spacer h-BN. Since the range of the eigenstates is $\ell_0$ is much longer we can use a local interaction defined by $U\equiv \int V_c(\rr)d^2r/\ell_0^2$. The average interaction is for occupied states in the flavors
$\alpha$ (including spin) with $i$ are states within each band, i.e. the eigenstates are now 16 long spinors $\varphi_{i\alpha}(\rr)$. 
\beq{18}
\langle{\cal H}\rangle_{int}=&&
\half U\ell_0^2\int d^2r\sum_{ij,\alpha\beta}[\varphi_{i\alpha}^\dagger(\rr)\varphi_{i\alpha}(\rr)\varphi_{j\beta}^\dagger(\rr)\varphi_{j\beta}(\rr)\nonumber\\&&-
\varphi_{j\beta}^\dagger(\rr)\varphi_{i\alpha}(\rr)\varphi_{i\alpha}^\dagger(\rr)\varphi_{j\beta}(\rr)]
\eeq

The first term is the direct (Hartree) interaction and the second one is the exchange (Fock) interaction. The spinor product implies that the exchange vanishes for $\alpha\neq\beta$ (orthogonal spinors). For $\alpha=\beta$ we note that within that subspace the eigenstates have the form $\tilde\varphi_i(\rr)b_i$ where $b_i$ are 2-spinors and $\tilde\varphi_i(\rr)$ are scalars; the latter represent the AA and AB Moiré lattice sites that are close on the scale $\ell_0$, hence they have a common $\rr$ dependence.
The disorder average involves, from the first Eq. \eqref{e17}, 
$\overline{|\tilde\varphi_i(\rr)|^2|\tilde\varphi_j(\rr)|^2}\sim \delta_{i,j}$ 
for both exchange and direct terms which therefore precisely cancel. 
The disorder average Hartree-Fock energy is then
\beq{19}
\overline{\langle{\cal H}_{int}\rangle}=\half U\ell_0^2\int d^2r\sum_{ij,\alpha\neq\beta}\overline{|\varphi_{i\alpha}|^2|\varphi_{j\beta}|^2}
\eeq

The initial interpretation \cite{ilani1} that has accounted for the main features of the cascade transitions, has assumed that there are four effective bands, each allowing filling of $-1<\nu<1$. We find (consistent with \cite{sachdev}), however, that there are 8 non-interacting flavors each allowing filling of $-\half<\nu<\half$. 

This apparent contradiction is reconciled by our key result that the states with opposite $v_m$ have vanishingly small interactions for all equal $v,s,\sign(E)$, as in Eq. (\ref{e12}b). In contrast, e.g. spin $\uparrow$ and $\downarrow$ states with the same $v_m,v,\sign(E)$ interact strongly as in Eq. (\ref{e12}a).  Our system is then equivalent to the Hartree-Fock result \eqref{e19} with 4 bands each having $-1<\nu< 1$. We note that each of the effective bands is separated into two by $\sign(E)$ of their kinetic energy. This separation is not due to the kinetic energy itself which is negligible, but due to the significantly different interaction terms as listed in \eqref{e17}.

Finally, we address the cascade transitions. The DOS divergence of each flavor implies that the occupied states are very near $E=0$ and therefore the kinetic energies are negligible.
Hence we need just the direct term with $\alpha\neq\beta$ in \eqref{e19} and use the interactions in Eq. \eqref{e17}. There are a number of choices for filling up the various flavor states, all leading to equivalent sequences of chemical potentials. In table \ref{table2} we have chosen states with equal $\uparrow,\,\downarrow$ populations since spin polarization (in the absence of magnetic fields) is not observed \cite{saito,ilani2}. We find that the stronger jumps are those into even fillings while the weaker jumps depend on the probed size $L$.\\

\section{Conclusions}

 Our first observation is the huge disorder parameter near the magic angle.  We claim that averaging on twist angle disorder is essential in measurements on scales $L\gg \ell_0$, where $\ell_0=0.2-0.5\,\mu$m is the disorder correlation length in TBG \cite{zeldov}, while less precisely known for TTG. 
Our second finding is a solution of a strong disorder problem with new results on $E\neq 0$ eigenstates of the Dirac equation with random gauge in terms of sum rules and their individual terms, Eqs. (\ref{e15}-\ref{e17}). Finally, we apply a Hartree-Fock average to show jumps of the chemical potential at integer $\nu$, as in
the cascade transitions \cite{ilani1,yacobi1,herrero,saito,yacobi2,liu}.

We recall that the shape of the cascade transitions is correlated with the measurement scale $L$  as discussed in the introduction. In particular the peaks are broad and shifted from integer $\nu$ when probing $\approx 0.5 \mu$m \cite{ilani1} or $\approx 0.1 \mu$ in TTG \cite{yacobi2}, while they are sharper and more accurately at integer $\nu$  when probing larger areas by either directly averaging over $\approx 2 \mu$m \cite{yacobi1} or by a global measurements over a few $\mu$m \cite{herrero,saito} and up to $\approx 10 \mu$m in TTG \cite{liu}.
 We propose then that the disorder average at $L\gg \ell_0$ leads to sharper transitions as in our Hartree-Fock result, accounting for the main feature of the data.

We finally note that the jump values in table II are independent of $\sigma$, i.e. the cascade transitions should not change with the distance to the magic angle, as long as $\sigma>2$. Upon increasing $|\theta-\theta_m|$ the disorder may reach $\sigma=2$, at this  elusive freezing transition we expect the cascade transitions to change considerably (see Eq. \eqref{e71}.


In conclusion, we have shown that twist angle disorder is crucial for interpreting data on long scales $L\gg\ell_0$ near the magic angle. This interpretation reveals novel insights on eigenstates of the Dirac equation in presence of a random gauge potential.\\

\acknowledgements 
We thank highly stimulating discussions with S. Ilani, E. Zeldov, E. Y. Andrei, A. Stern, F. Guinea, T. Giamarchi, D. Bernard and E. Akkermans.
PLD thanks Ben-Gurion University Physics department for hospitality. B. H. thanks the Laboratoire de Physique de l'Ecole Normale Sup\'erieure for hospitality

\begin{widetext}
	
\appendix

\section{Symmetries and the reduced Hamiltonian}

In this appendix we derive the symmetries of the effective Hamiltonian in presence of a random vector potential and derive the reduced Hamiltonian near the magic angle for both TBG and TTG.

A given Dirac point in graphene, i.e. a given valley, is split by the relative rotation $\theta$ into two Dirac points that are connected by the wavevecors \cite{BM}
\beq{31}
\qq_1=k_\theta (0,\,-1),\qquad \qq_2=\half k_\theta (\sqrt{3},\,1),\qquad \qq_3=\half k_\theta(-\sqrt{3},\,1)
\eeq
where $k_\theta= \frac{4\pi}{3L_m}$. The effective Hamiltonian near these Dirac points is 
\beq{32}
&&{\cal H}^{(1)}=\begin{pmatrix}v_F{\bm \sigma}\cdot[-i{\bm\nabla}_\rr -\tilde\AB(\rr)] & \eexp{i(\theta/4)\sigma_z}\hat T(\rr)
	\eexp{i(\theta/4)\sigma_z} \\
	\eexp{-i(\theta/4)\sigma_z}\hat T^\dagger(\rr)\eexp{-i(\theta/4)\sigma_z} & v_F{\bm \sigma}\cdot[-i{\bm\nabla}_\rr+\tilde\AB(\rr)] \end{pmatrix}\nonumber\\&&
\hat T(\rr)=\sum_{i=1}^3 \eexp{i\qq_i\cdot\rr}T_i,\,\,\, T_j=u_0\mathbbm{1} +u_1\cos[(j-1)\phi]\sigma_x+u_1\sin[(j-1)\phi]\sigma_y,\qquad j=1,2,3
\eeq
where $\phi=2\pi/3$, $v_F$ is the bare Dirac velocity of a single layer,  $\sigma_{x,y,z}$ are Pauli matrices and the interlayer couplings at either AA or AB overlaps are $u_0$ and $u_1$, respectively.
This effective Hamiltonian with $\tilde \AB=0$ was given in Ref. \onlinecite{BM} and was extended \cite{paco2,mele} to cases with a vector potential induced by lattice relaxation. In our case $\tilde \AB(\rr)$ is random, it has opposite signs on the two layers since $\tilde \AB(\rr)\sim \pm\delta\theta$.

The Hamiltonian near the second valley is
\beq{33}
{\cal H}^{(2)}=\begin{pmatrix}{v_F\bm \sigma}^*\cdot[i{\bm\nabla}_\rr-\tilde\AB(-\rr)] & \eexp{-i(\theta/4)\sigma_z}\hat T^*(\rr)
	\eexp{-i(\theta/4)\sigma_z} \\
	\eexp{i(\theta/4)\sigma_z}\hat T^{*\dagger}(\rr)\eexp{i(\theta/4)\sigma_z} & v_F{\bm \sigma}^*\cdot[i{\bm\nabla}_\rr+\tilde\AB(-\rr)] \end{pmatrix}
\eeq
This is obtained by rotating \eqref{e32} by $C_{2z}$ which involves the operator $R=\{\rr\rightarrow -\rr\}$ and interchange carbon sites $A\leftrightarrow B$, i.e. $\sigma_x$, hence
${\cal H}^{(2)}=R\sigma_x{\cal H}^{(1)}\sigma_x R$
using $\sigma_x T(-\rr)\sigma_x=T^*(\rr)$.

We note that the factors $\eexp{\pm i\theta\sigma_z/4}$  in Eqs. (\ref{e32},\ref{e33}) are commonly neglected \cite{paco2,mele,bernevig} since $\theta$ is small. However, these factors lead to non-abelian disorder that can be relevant \cite{fosterW}. These factors modify the $u_0$ interlayer coupling to $u_0\eexp{i\theta\sigma_z/2}$ so that Eqs. (\ref{e32},\ref{e33}) contain to leading order to ${\cal H}'=\mp\half u_0\theta\sigma_z\eta_y\sum_j\eexp{i\qq_j\cdot\rr}$ where $\eta_{x,y,z}$ are Pauli matrices in the subspace of the two layers. Since $\theta(\rr)=\overline\theta +\delta\theta(\rr)$ ${\cal H}'$ represents non-abelian disorder. In comparison, the abelian disorder $v_F{\bm\sigma}\cdot\tilde\AB(\rr)\eta_z$ contains $\approx v_F\delta\theta \beta/a'$ [Eq. (3)], hence the ratio of coefficients is $\frac{u_0/2}{v_F\beta/a'}\approx 10^{-2}$, assuming $u_0\approx u_1\approx \frac{4\pi v_F\theta}{3a\sqrt{3}}$ near the magic angle $\theta=1.1^\circ\approx 0.02$rad. In the following we therefore neglect these $\eexp{\pm i\theta\sigma_z/4}$ factors.

We consider next the symmetries of the Hamiltonian, extending previous studies \cite{sheffer,bernevig} to our case with a random $\tilde\AB(\rr)$.
Note first that the Hamiltonian of both layers, with Pauli matrices ${\bm\tau}$ in the valley subspace possess the symmetry operation $\sigma_x\tau_x R$
\beq{34}
\sigma_x\tau_x R {\cal H} \sigma_x\tau_x R=R\begin{pmatrix} 0 & \sigma_x\\ \sigma_x & 0 \end{pmatrix}\begin{pmatrix} {\cal H}^{(1)} & 0 \\
	0 & {\cal H}^{(2)} \end{pmatrix}\begin{pmatrix} 0 & \sigma_x\\ \sigma_x & 0 \end{pmatrix}R={\cal H}
\eeq
This is an extension of the single valley symmetry \cite{sheffer} $\sigma_x {\cal K} R$ for the case $\tilde\AB(\rr)=0$, where $\cal K$ represents complex conjugation. 

Since we neglect the small $\theta$ in Eqs. (\ref{e32},\ref{e33}) the system has a particle-hole (p-h) symmetry operator ${\cal C}=\eta_y\sigma_x \cal K$,
\beq{35}
{\cal C}{\cal H}^{(1)}{\cal C}={\cal K}\begin{pmatrix} 0 & -i\sigma_x\\ i\sigma_x & 0 \end{pmatrix}\begin{pmatrix} {\bm\sigma}\cdot[-i{\bm\nabla}-\tilde\AB(\rr)] & \hat T (\rr)\\ \hat T ^\dagger (\rr) &  {\bm\sigma}\cdot[-i{\bm\nabla}+\tilde\AB(\rr)]\end{pmatrix}\begin{pmatrix} 0 & -i\sigma_x\\ i\sigma_x & 0 \end{pmatrix} {\cal K}=-{\cal H}^{(1)}\nonumber\\
\eeq
and similarly with ${\cal H}^{(2)}$. This is actually the same operator as in the $\tilde A(\rr)=0$ case \cite{sheffer}.
Thus, for each eigenstate $|\psi\rangle$ with eigenvalue $E$ there is an eigenstate ${\cal C}|\psi\rangle$ with eigenvalue $-E$. In view of the symmetry \eqref{e34} the p-h symmetry can also be represented by $\tau_x\eta_y R\cal K$, that mixes the two valleys.  
We note that a time reversal operator $\tau_x{\cal K}$, which is a valid symmetry for $\tilde\AB(\rr)=0$, is broken when  $\tilde\AB(\rr)\neq 0$. The symmetries in the reduced Hamiltonian near the magic angle are also considered in section II.

We proceed to find the reduced Hamiltonian near the magic angle. We follow the perturbation expansion \cite{BM,paco2,mele} in $u_0,\,u_1$ connecting a state in layer I to 3 states layer II .
We then find an effective Hamiltonian near the Dirac point of layer I. The one-layer Hamiltonian $v_F{\bm\sigma}\cdot[-i{\bm\nabla}-\tilde\AB(\rr)]$ with a state $\psi_0(\rr)$ is connected by $T_j$ to the 3 states $\psi_j(\rr)\eexp{-i\qq_j\cdot\rr}$, $j=1,2,3$. Ignoring the small $\theta$ rotation (exact for $u_0=0$) the 4x4 Hamiltonian Eq. \eqref{e32} becomes a 8x8 one, 
\beq{36}
&&\tilde{\cal H}=v_F\begin{pmatrix} {\bm\sigma}\cdot[-i{\bm\nabla}-\tilde\AB(\rr)]  & T_1/v_F & T_2/v_F & T_3/v_F \\ T_1/v_F & {\bm\sigma}\cdot[-i{\bm\nabla}+\tilde\AB(\rr)-\qq_1] & 0 & 0 \\ T_2/v_F & 0 &  {\bm\sigma}\cdot[-i{\bm\nabla}+\tilde\AB(\rr)-\qq_2] & 0 \\ T_3/v_F & 0 & 0 &  {\bm\sigma}\cdot[-i{\bm\nabla}+\tilde\AB(\rr)-\qq_3] \end{pmatrix}
\eeq
Note that $\tilde A(\rr)$ is slowly varying on the scale $L_m$ so that it is kept on the diagonal. This is in contrast to the vector potentials from lattice relaxation \cite{paco2,mele} that carry momenta $\qq_i-\qq_j$ ($i\neq j$) and therefore appear in this matrix in off-diagonal positions; the latter case results in a shift of the magic angle \cite{paco2,mele}.

We next repeat the known limit $i{\bm\nabla}=\tilde\AB(\rr)=0$ showing that it has an eigenstate $\Psi_0=(\psi_0, \,\psi_1,\,\psi_2,\,\psi_3)$ with zero eigenvalue. Defining matrices $ h_j, \, j=1,2,3$, 
\beq{37}
h_j= && -v_F{\bm \sigma}\cdot\qq_j= iv_Fk_\theta \begin{pmatrix} 0 & -\eexp{-i(j-1)\phi} \\ \eexp{i(j-1)\phi} & 0 \end{pmatrix}\nonumber\\
h_j^{-1}=&& h_j/(v_Fk_\theta) ^2,\qquad T_jh_j^{-1}T_j=(u_0^2-u_1^2)h_j/(v_Fk_\theta)^2
\eeq 
Note that $\sum_j T_jh_j^{-1}T_j=0$ for any $u_0,\,u_1$. The eigenvalue equation of \eqref{e36} yields
\beq{38}
\text {lines 2,3,4}\qquad & T_j\psi_0+h_j\psi_j=0\qquad \Rightarrow \psi_j=-h_j^{-1}T_j\psi_0\nonumber\\
\text {line 1}\qquad &
\sum_jT_j\psi_j=(\sum_jT_jh_i^{-1} T_j)\psi_0=0
\eeq
which identifies $\psi_j$ and proves that it has vanishing eigenvalue. Assuming $\langle \psi_0|\psi_0\rangle=1$ we find the normalization $\langle\Psi_0 |\Psi_0\rangle=1+3(\bar u_0^2+\bar u_1^2)$
using $(h_j^{-1})^2=\mathbbm{1}/(v_Fk_\theta)^2$ and defining $\bar u_i=u_i/v_Fk_\theta,\, i=1,2$.

Consider next the effective Hamiltonian projected to the $\psi_0$ space, to 1st order in $-i{\bm\nabla},\, \tilde\AB(\rr)$, defined by
\beq{39}
\langle\psi_0|{\cal H}_{eff}|\psi_0\rangle \sim\langle \Psi_0+O(k,\tilde\AB)|v_F{\bm \sigma}\cdot
[-i{\bm\nabla}\mathbbm{1}-\tilde\AB(\rr)D]|\Psi_0+O(k,\tilde\AB)\rangle
\eeq
where $D$ is a diagonal matrix with elements (1,-1,-1,-1). 
Including the normalization we obtain
\beq{40}
\langle\psi_0|{\cal H}_{eff}|\psi_0\rangle &&=\frac{1}{1+3(\bar u_0^2+\bar u_1^2)}\langle \Psi_0|v_F{\bm \sigma}\cdot
[\kk\mathbbm{1}-\tilde\AB(\rr)D]|\Psi_0\rangle\nonumber\\&&
=\frac{v_F}{1+3(\bar u_0^2+\bar u_1^2)}
\langle \psi_0|{\bm \sigma}\cdot [\kk-\tilde\AB(\rr)]+\sum_j T_j(h_j^{-1})^\dagger
{\bm \sigma}\cdot [\kk+\tilde\AB(\rr)]h_j^{-1}T_j |\psi_0\rangle\nonumber\\&&
=\frac{v_F}{1+3(\bar u_0^2+\bar u_1^2)}\langle \psi_0|{\bm \sigma}\cdot [\kk-\tilde\AB(\rr)]-3\bar u_1^2 {\bm \sigma}\cdot [\kk+\tilde\AB(\rr)]|\psi_0\rangle
\eeq 
This then identifies the reduced Hamiltonian, Eq. \eqref{e05}.

We note that in addition to $\tilde\AB(\rr)\approx \frac{\beta}{a'}\delta\theta$ (Eq. (3)) the Dirac Hamiltonian Eq. (5) contains a $\delta\theta$ dependence from $v(\theta+\delta\theta)=v(\theta)+v_F\frac{\delta\theta}{\theta}$, hence a Hamiltonian term $\approx -iv_F\frac{\delta\theta(\rr)}{\theta}{\bm\sigma}\cdot{\bm\nabla}$, possibly relating to interference between different Moiré periodicities \cite{foster3}. However, the ratio of this term to that of $\tilde\AB(\rr)$ is small $\frac{a'}{\ell_0\theta\beta}\ll 1$ and is therefore neglected, taking $|\bm\nabla|\sim 1/\ell_0$.

We next interchange the two layers with an operator $\eta_x$, i.e.
\beq{41}
\eta_x{\cal H}^{(1)}\eta_x=\begin{pmatrix}{\bm \sigma}\cdot[-i{\bm\nabla}_\rr +\tilde\AB(\rr)] & \eexp{-i(\theta/4)\sigma_z}\hat T^\dagger(\rr)
	\eexp{-i(\theta/4)\sigma_z} \\
	\eexp{i(\theta/4)\sigma_z}\hat T(\rr)\eexp{i(\theta/4)\sigma_z} & {\bm \sigma}\cdot[-i{\bm\nabla}_\rr-\tilde\AB(\rr)] \end{pmatrix}
\eeq
and build the perturbation expansion with a state on layer II that is coupled to 3 states in layer I. This corresponds to expanding near the Dirac point of layer II representing a distinct Moir\'e valley.
Neglecting $\theta$ the only change in the perturbative solution is $\qq_j\rightarrow -\qq_j$ which does not change the result.
The effective Hamiltonian is then the same as in \eqref{e40}, i.e. Eq. (5), except significantly that $\tilde\AB(\rr)\rightarrow -\tilde\AB(\rr)$.  Hence we represent $\eta_x$ explicitly by $\tilde\AB(\rr)\rightarrow -\tilde\AB(\rr)$.

Considering next the other graphene valley with ${\cal H}^{(2)}$, the perturbation proceeds with $h_j\rightarrow -h_j^*=-h_j(-\phi)$, so the effective Hamiltonian is \eqref{e40} except $\tilde\AB(\rr)\rightarrow \tilde\AB(-\rr)$ and ${\cal H}_{eff}\rightarrow {\cal H}_{eff}^*$.

The system is thus decoupled into three subspaces labelled with $s,\, v^m,\, v$ where $s=\pm$ for spin, $v^m=\pm$ for Moiré valley and $v=\pm$ for graphene valley. Each of the resulting 8 bands has a single Dirac cone (positive and negative energies) with two spinor components defined (in real space) on AA and AB sites, respectively.

{\it Trilayer twisted graphene:} We finally consider mirror symmetric TTG, in particular the ABA configuration, as in the relevant experiments \cite{yacobi2,liu}.  We also note that the ABA configuration is energetically favored, having a stronger interlayer coupling \cite{kaxiras3}. 

We derive the reduced Hamiltonian, following the previous TBG derivation and its TTG extension \cite{macdonald}. We allow initially for arbitrary variations in the local twist angle of each layer, leading to 3 random vector potentials $\AB_j(\rr),\,j=1,2,3$.
The previous interlayer matrices $T_j$ for $AA$ overlap at the middle of the Moiré unit cell need to be replaced by 
$T_j\rightarrow \eexp{-i\qq_j\cdot\rr_0}T_j$ for $AB$ overlap, where $\rr_0=\frac{4\pi}{3\sqrt{3}k_\theta}(-1,0)$; eventually this phase factor has no effect. The Hamiltonian \eqref{e36} becomes for the ABA configuration 
\beq{203}
&&\tilde{\cal H}={\scriptsize \begin{pmatrix} v_F{\bm\sigma}\cdot[-i{\bm\nabla}-\AB_1(\rr)]  & T_1 & T_2 & T_3 & 0 \\ T_1^\dagger & v_F{\bm\sigma}\cdot[-i{\bm\nabla}-\AB_2(\rr)-\qq_1] & 0 & 0 & T_1^\dagger \\ T_2^\dagger & 0 &  v_F{\bm\sigma}\cdot[-i{\bm\nabla}-\AB_2(\rr)-\qq_2] & 0 & T_2^\dagger \\ T_3^\dagger & 0 & 0 &  v_F{\bm\sigma}\cdot[-i{\bm\nabla}-\AB_2(\rr)-\qq_3] & T_3^\dagger \\ 0 & T_1 & T_2 & T_3 & v_F{\bm\sigma}\cdot[-i{\bm\nabla}-\AB_3(\rr)] \end{pmatrix}}\nonumber\\
\eeq
At $i{\bm\nabla}=\AB_j(\rr)=0$ the eigenstates
$\Psi_0=(\alpha,\beta_1,\beta_2,\beta_3,\gamma)$ for vanishing eigenvalue satisfy
\beq{204}
\text {lines 2,3,4}\qquad & T_j^\dagger\alpha+h_j\beta_j+T_j^\dagger\gamma=0\qquad \Rightarrow \beta_j=-h_j^{-1}T_j^\dagger (\alpha+\gamma)\nonumber\\
\text {lines 1,5}\qquad &
\sum_jT_j\beta_j=0=-\sum_jT_jh_j^{-1}T_j^\dagger (\alpha+\gamma)
\eeq
and the mirror symmetry allows $\alpha=\pm\gamma$. The normalization for the case $\alpha=\gamma$ is
\[\langle\Psi_0|\Psi_0\rangle=2+\langle \alpha+\gamma |\sum_j T_j (h_j^{-1})^\dagger h_j^{-1}T_j^\dagger |\alpha+\gamma\rangle =2+12(\bar u_0^2+\bar u_1^2)\equiv C\]
The effective Hamiltonian, projected on $|\alpha\rangle$, to first order in $i{\bm\nabla},\,\AB_j(\rr)$ is
\beq{205}
&&\langle\alpha |{\cal H}_{eff}|\alpha\rangle =\langle\Psi_0| v_F{\bm\sigma}\cdot {\scriptsize\begin{pmatrix} [-i{\bm\nabla}-\AB_1(\rr) & 0 & 0 & 0 & 0 \\ 0 & -i{\bm\nabla}-\AB_2(\rr) & 0 & 0 & 0 \\ 0 & 0 & -i{\bm\nabla}-\AB_2(\rr) & 0 & 0 \\ 0 & 0 & 0 & -i{\bm\nabla}-\AB_2(\rr) & 0 \\ 0 & 0 & 0 & 0 & -i{\bm\nabla}-\AB_3(\rr)\end{pmatrix} }|\Psi_0\rangle \nonumber \\
=&&\frac{v_F}{C}\langle\alpha|{\bm\sigma}\cdot [-2i{\bm\nabla}-\AB_1(\rr)-\AB_3(\rr)]+\sum_j\langle \alpha+\gamma |T_j(h_j^{-1})^\dagger {\bm\sigma}[-i{\bm\nabla}-\AB_2(\rr)]h_j^{-1}T_j^\dagger |\alpha+\gamma\rangle \nonumber\\
&& =\frac{v_F}{C}\langle\alpha | {\bm\sigma}\cdot [-2i{\bm\nabla}-\AB_1(\rr)-\AB_3(\rr)]-12\bar u_1^2{\bm\sigma}[-i{\bm\nabla}-\AB_2(\rr)]|\alpha\rangle\nonumber\\
&& \Rightarrow\,\, {\cal H}_{eff}=\frac{v_F}{C}{\bm \sigma}\cdot [-2i{\bm\nabla}(1-6\bar u_1^2)-(\AB_1(\rr)+\AB_3(\rr)-12\bar u_1^2 \AB_2(\rr))] 
\eeq
The Dirac velocity is then
\beq{206}
v=v_F\frac{1-6\bar u_1^2}{1+6(\bar u_0^2 +\bar u_1^2)}
\eeq
The critical value where $v$ vanishes, relative to that of TBG, is $\bar u_1^c(TTG)=\bar u_1^c(TBG)/\sqrt{2}$; since $\bar u_1\sim 1/\theta$ the magic angle satisfies $\theta_m(TTG)=\sqrt{2}\theta_m(TBG)$.
The disorder terms can be written as
\[\frac{v}{4}\{\frac{1+6\bar u_1^2}{1-6\bar u_1^2}[\AB_1(\rr)+\AB_3(\rr)-2\AB_2(\rr)]+\AB_1(\rr)+\AB_3(\rr)+2\AB_2(\rr)\} \]
The first term diverges at the critical point, hence dominant, while the second term is finite and is therefore neglected. The effective disorder vector potential is then
\beq{207}
&&\AB(\rr)=\frac{1+6\bar u_1^2}{4(1-6\bar u_1^2)}[\AB_1(\rr)+\AB_3(\rr)-2\AB_2(\rr)]\nonumber\\&&
{\cal H}_{eff}=v{\bm\sigma}\cdot[-i{\bm\nabla}-\AB(\rr)]
\eeq
exactly the same form as Eq. \eqref{e05}. 
In terms of the local twist angles $\delta\theta_i(\rr)$ we have 
$\AB(\rr)\sim \delta\theta_1(\rr)+\delta\theta_3(\rr)-2\delta\theta_3(\rr)=\delta\theta_{12}(\rr)+\delta\theta_{32}(\rr)$. We note that for the solution $\alpha=-\gamma$ Eq. \eqref{e204} implies $\beta_j=0$, hence the interlayer coupling has no effect and the Dirac Hamiltonian is just $v_F{\bm\sigma}\cdot[-i{\bm\nabla}-\AB_i(\rr)],\,i=1,3$, i.e. the velocity $v_F$ is maintained and the disorder has a minor effect.

\section{Density verlaps for $E=0$ states}

We wish to evaluate the density overlaps between the various $E=0$ eigenstates listed in table I. By density overlap we mean the disorder average $\overline{|\varphi_i(\rr)|^2|\varphi_j(\rr)|^2}$ for various 2-spinor eigenstates of the Dirac equation with a random gauge potential.

We start with replica symmetric method that applies to weak disorder \cite{ludwig}. Consider, where eventually $n\rightarrow 0$, and using $\langle V^2(\rr)\rangle=\half\sigma\ln\frac{L}{\ell_0},\,
\langle V(\rr)V(0)\rangle=\half\sigma\ln\frac{L}{r}$, 
\beq{51}
&&P_{q,q'}=\overline{ \frac{\int \eexp{2q'V(\rr)}d^2r}{[\int \eexp{2V(\rr)}d^2r]^q}}=
\overline{\int\eexp{2q'V(\rr)}d^2r\,\Pi_{j=1}^{n-q}\int d^2r_j \eexp{2V(\rr_j)}}
=\int d^2rd^2r_1...d^2r_{n-q}\eexp{2\overline{[q'V(\rr)+\sum_{j=1}^{n-q}V(\rr_j)]^2}}
\nonumber\\&& =\eexp{\sigma (q'^2+n-q)\ln\frac{L}{\ell_0}}\int d^2rd^2r_1...d^2r_{n-q}\eexp{2\sigma q'\sum_{j=1}^{n-q}\ln\frac{L}{|\rr-\rr_j|}
	+2\sigma\sum_{i<j}^{n-q}\ln\frac{L}{|\rr_i-\rr_j|}}\nonumber\\
&&\approx \left(\frac{L}{\ell_0}\right)^{\sigma (q'^2+n-q)}L^{2+2(n-q)}
\, \overset{n\rightarrow 0}{=}\, \left(\frac{L}{\ell_0}\right)^{\sigma (q'^2-q)}L^{2-2q }
\eeq
Since $n \to 0$ at the end, we always omit factors $Z(1)^n$ in denominators, where $Z(1)=\int \eexp{2  V(\rr)}d^2r$.
The last step assumes that all differences $|\rr-\rr_i|,\,|\rr_i-\rr_j|\sim L$ so the the dimension of the integral is $2+2(n-q)$, implying that $\sigma$ is not too large.
For the needed intensity overlaps $P_{2,2}\sim L^{2\sigma-2}$ and $P_{2,0}\sim L^{-2-2\sigma}$.

To treat large $\sigma$ we follow the packet partition of replicas leading to replica symmetry breaking (RSB), as discussed in Ref. \cite{scheidl} appendix D and in Ref. \cite{carpentier2} Eqs. 77-81. For strong $\sigma$ we expect strong attraction between replica variables, thus we allow the option of forming packets where $p$ replica variables $\rr_i$ are close to each other, eventually $p$ is determined by a minimum condition.
Consider first the packet division of $P_{q0}$. In \eqref{e51} divide the $n-q$ replicas into $(n-q)/p$ packets having each $p$ replicas. Each packet has $\rr_{ki}\,,k=1,...,p$ close to each other while the inter-packet distance is $\sim L$. Thus the number of integrations $\sim L^2$ is $1+(n-q)/p$ (including 1 from the nominator of $P_{q0}$),
\beq{52}
P_{q,0}=&&\eexp{2\sigma (n-q)\ln\frac{L}{\ell_0}}L^{2+2\frac{n-q}{p}}\{\int\Pi_{i=1}^{(n-q)/p}\,\, \frac{d^2r_{1i}...d^2r_{pi}}{L^2}
\eexp{2\sigma\sum_{k,k'=1}^p\sum_{i<j}^{(n-q)/p}\ln\frac{L}{|\rr_{ki}-\rr_{k'j}|}}\}\eexp{2\sigma \half p(p-1)\frac{n-q}{p}\ln\frac{L}{\ell_0}} \nonumber\\&&
\approx   L^{2+2\frac{n-q}{p}}   (\frac{L}{\ell_0})^{\sigma(n-q)+\sigma(p-1)(n-q)}\cdot 
\ell_0^{2(p-1)\frac{n-q}{p}}\overset{n\rightarrow 0}{=}
\ell_0^{2-2q}(\frac{L}{\ell_0})^{2-2q(\frac{1}{p}+\half\sigma p)}
\eeq
where the \{\} integral is by assumption convergent for each packet so it is $\approx \ell_0^{2(p-1)}$.
Minimizing 
$\frac{1}{p}+\half\sigma p$ leads to $p=\min[\frac{1}{\sqrt{\sigma/2}},1]$, hence a critical $\sigma_c=2$. For $\sigma<2$ there is no RSB ($p=1$) and the result coincides with \eqref{e51}. For $\sigma>2$ RSB is present ($p<1$), hence
\beq{53}
P_{q,0}\approx \ell_0^{2-2q} (\frac{L}{\ell_0})^{2-4q\sqrt{\sigma/2}} \qquad \sigma>2\,.
\eeq

In the following we actualy need a slightly different average  
\beq{54}
P'_{2,0}=\overline{\frac{\int d^2r}{\int d^2r\,\eexp{2V(\rr)}\int d^2r\,\eexp{-2V(\rr)}}}
=\int d^2r \overset{n-1}{\underset{i=1}{\Pi}}
d^2r_id^2r'_i\,\overline{ \eexp{2\sum_{i=1}^{n-1}[V(\rr_i)-V(\rr'_i)]}}
\eeq
Since the groups $\rr_i$ and $\rr'_i$ come with opposite signs of $V$ they interact with $-\sigma<0$ which does not favor packet formation. Hence we treat the two groups separately. Divide each group of $n-1$ replicas into $i=1,2,...,(n-1)/p$ packets having each $p$ replicas. Each packet has either $\rr_{ki}\,,k=1,...,p$ or $\rr'_{ki}\,,k=1,...,p$  close to each other while the inter-packet distance is $\sim L$. Thus the number of integrations $\sim L^2$ is $1+2(n-1)/p$  
\beq{55}
&&P'_{20}=\eexp{2\sigma(n-1)\ln\frac{L}{\ell_0}}L^{2+4(n-1)/p}
\{\int\Pi_{i=1}^{(n-1)/p}\,\, \frac{d^2r_{1i}...d^2r_{pi}}{L^2}\int\Pi_{i=1}^{(n-1)/p}\,\, \frac{d^2r'_{1i}...d^2r'_{pi}}{L^2}\nonumber\\&& \eexp{[2\sigma\sum_{k,k'=1}^p\sum_{i<j}^{(n-1)/p}\ln\frac{L}{|\rr_{ki}-\rr_{k'j}|}]+[\text{all}\,\, \rr_{ki}-\rr_{k'j}\rightarrow \rr'_{ki}-\rr'_{k'j}]- 2 [\text{all}\,\, \rr_{ki}-\rr_{k'j}\rightarrow \rr_{ki}-\rr'_{k'j}]}\} \eexp{ 2\sigma\ln\frac{L}{\ell_0}\half p(p-1)\cdot 2(n-1)/p}\nonumber\\&&
\approx L^{2+4(n-1)/p}(\frac{L}{\ell_0})^{2\sigma(n-1)+2\sigma(p-1)(n-1)}
\ell_0^{2(p-1)2(n-1)/p}\,\overset{n\rightarrow 0}{=}\,\ell_0^{-2}(\frac{L}{\ell_0})^{2-4(\frac{1}{p}+\half\sigma p)}
\eeq
where $\{...\}$ is $L$ independent. The result is identical to $P_{2,0}$.

For $q'\neq 0$ the $\rr,\,\rr_i$ integrations are not symmetric, therefore allow for one packet of size $p_0$ that interacts with $q'V(\rr)$, i.e. its variables are $\rr,\rr_{k0}$, $k=1,\dots,p_0$, the other variables are divided to packets of size $p$, i.e. $i=1,...,\frac{n-q-p_0}{p}$ packets. Each packet has a c.m. that spans the area $L^2$ while the spacing between packet members is $\approx \ell_0$. Each intrapacket interaction is $\sim\ln\frac{L}{\ell_0}$ while all interpacket terms are 
$\sim \ln\frac{L}{|\rr_{ki}-\rr_{k'j}|},\,i\neq j$. Hence,
\beq{56}
&&P_{q,q'}=\eexp{\sigma (q'^2+n-q)\ln\frac{L}{\ell_0}} L^{2+2\frac{n-q-p_0}{p}}\{
\int\frac{d^2r}{L^2} \overset{p_0}{\underset{k=1}{\Pi}}
\int d^2r_{k0}
\overset{n-q-p_0}{\underset{i=1}{\Pi}}
\frac{1}{L^2} 
\overset{p}{\underset{k=1}{\Pi}}
\int d^2r_{ki}
\nonumber\\&&\eexp{2\sigma q'\sum_{k=1}^p\sum_{j=1}\ln\frac{L}{\rr-\rr_{kj}|}+
	2\sigma\sum_{k=1}^{p_0}\sum_{k'=1}^p\sum_j\ln\frac{L}{|\rr_{k0}-\rr_{k'j}|}
	+2\sigma\sum_{k,k'=1}^p\sum_{i<j}\ln\frac{L}{|\rr_{ki}-\rr_{k'j}|}}\}
\nonumber\\&& \eexp{2\sigma \ln\frac{L}{\ell_0}[ q'p_0 +\half p_0(p_0-1)+
	\half p(p-1)\frac{n-q-p_0}{p}]}
\eeq
The $\{...\}$ factor is again $L$ independent, hence
\beq{57}
\frac{\ln P_{q,q'}}{\ln L}=&&\sigma (q'^2+n-q)+2(1+\frac{n-q-p_0}{p})+2\sigma q'p_0+\sigma p_0(p_0-1)+\sigma p(p-1)\frac{n-q-p_0}{p}\nonumber\\&&
\overset{n\rightarrow 0}{=}\sigma q'^2+2-2(\frac{1}{p}+\half\sigma p)(q+p_0)+
2\sigma q'p_0+\sigma p_0^2 
\eeq

Optimizing with respect to $p$ leads as before to $p= \min(1, 1/\sqrt{\sigma/2})$, hence a critical $\sigma_c=2$. Optimizing with respect to $p_0$ yields three regimes 
\beq{58}
p_0 + q' = \begin{cases} 
	q' \quad , \quad \sigma < \frac{2}{2 q'-1}  \\
	\frac{1}{2} (1 + \frac{2}{\sigma})  \quad , \quad \frac{2}{2 q'-1} < \sigma<2 \\
	1/\sqrt{\sigma/2}  \quad , \quad \sigma>2 
\end{cases} 
\eeq
In the first two regimes we assume $q' \geq 1$.
In the first, small $\sigma <  \frac{2}{2 q'-1} $ regime one has $p=1,\,p_0=0$ which leads to 
$P_{q,q'}\sim L^{\sigma(q'^2-q)+2-2q}$ in agreement with \eqref{e01},
while in the third, $\sigma>2$ regime one finds
\beq{59}
P_{q,q'}\approx \left(\frac{L}{\ell_0}\right)^{-4\sqrt{\sigma/2}(q-q')}\ell_0^{2-2q},\qquad \sigma>2,\,\, q'>0
\eeq
Note that at $q'= 0$ this result differs from the previous $P_{q,0}$ in \eqref{e02} by a factor $L^2$, in this case the variable $p_0$ is redundant since there is no $q'V(\rr)$ to induce it.

Finally, we find an intermediate regime for $\frac{2}{2 q'-1} < \sigma<1$.
This regime was obtained previously in the case $q'=q$ by a
completely different replica method, and called "pre-freezing" \cite{fyodorov}
(in that case, it represents the disorder-averaged inverse participation ratio).
Here we find in that regime
\beq{60}
P_{q,q'}\approx \left(\frac{L}{\ell_0}\right)^{-(2+ \sigma)(q-q') - \frac{(2-\sigma)^2}{ 4 \sigma} }  \ell_0^{2-2q},\qquad \frac{2}{2 q'-1} < \sigma<2,\,\, q'>1
\eeq
Although it is gratifying to obtain here the pre-freezing regime by a distinct replica method,
we will not need it for our applications which focus on large $\sigma$.
\\

{\bf Remark}. It is important to note that the $P_{q,q'}$ computed here are
{\it disorder averages} of the ratios $\overline{Z(q')/Z(1)^q}$, where we denote
$Z(q)=\int \eexp{2  q V(\rr)}d^2r$, also called annealed quantities. One can also
compute the {\it typical values} of these ratios, as was done for $q'=q$
in Ref. \cite{carpentier2} Eq. 138, 
as well as in Ref. \cite{chamon} Eq. 4.5 within Derrida's REM model.
These studies have shown that at large $L$
\beq{61}
\frac{\log Z(q)}{\log L} \to \begin{cases} 2  +  \sigma q^2 \quad , \quad \sigma < 2/q^2 \\
	4 q \sqrt{\sigma/2} \quad , \quad \sigma > 2/q^2 \end{cases} 
\eeq 
Since $\log(Z(q')/Z(1)^q) = \log Z(q') - q \log Z(1)$ it leads to three regimes 
\beq{62}
\frac{\log(Z(q')/Z(1)^q)}{\log L} \to 
\begin{cases}  
	\sigma ( (q')^2 - q) + 2 -  2 q \quad , \quad \sigma < 2/(q')^2 \\
	4 q' \sqrt{\sigma/2} - q (2+\sigma) \quad , \quad 2/(q')^2 < \sigma < 2 \\
	4 \sqrt{\sigma/2} ( q' - q)  \quad , \quad  \sigma>2
\end{cases} 
\eeq 
Comparing with the above we see that the typical and average differ only
in the intermediate disorder regime, i.e. the nature of the pre-freezing is
slightly different in both cases.

Consider next the density overlap
\beq{63}
Q_{1,-1}=\overline{ \frac{\int d^2r\, \eexp{2V(\rr)-2V(-\rr)}}{\int d^2r\,\eexp{2V(\rr)}\int d^2r\,\eexp{-2V(-\rr)}}}
=\int d^2r
\overset{n-1}{\underset{i=1}{\Pi}}
d^2r_id^2r'_i\,\overline{ \eexp{2V(\rr)-2V(-\rr)+2\sum_{i=1}^{n-1}[V(\rr_i)-V(-\rr'_i)]}}
\eeq
Since the groups $\rr,\rr_i$ and $-\rr,-\rr'_i$ come with opposite signs of $V$ they interact with $-\sigma<0$ which does not favor packet formation. Hence we treat the two groups separately. Consider $\rr_{k1},\,k=1,...,p_1$ as one packet with $p_1-1$ replicas, that includes $\rr=\rr_{11}$ for convenience, and then packets $\rr_{ki},\,k=1,...,p$ each with $p$ replicas and $i=2,3,...,n'=\frac{n-p_1}{p}+1$, all together represent $n-1$ replicas. Similarly consider all $\rr_{ki}\rightarrow -\rr'_{ki}$ with $\rr'_{11}=-\rr$. Note that the pair $\rr, -\rr$ cannot be in the same packet since then $r\lesssim\ell_0$ and the c.m. of the packet cannot span the whole space, hence the packets $\rr_{k1},\,\rr'_{k1}$ are distinct. Thus 
\beq{64}
&&Q_{1,-1}=\eexp{\sigma\ln\frac{L}{\ell_0}(2+2n-2)}L^{2+2(n'-1)\cdot 2}
\{\int\frac{d^2r}{L^2}
\overset{p_1}{\underset{k=2}{\Pi}}
\int d^2r_{k1}
\overset{n'}{\underset{i=2}{\Pi}}
\frac{1}{L^2}
\overset{p}{\underset{k=1}{\Pi}}
d^2r_{ki}\cdot
\overset{p_1}{\underset{k=2}{\Pi}}
\int d^2r'_{k1}
\overset{n'}{\underset{i=2}{\Pi}}
\frac{1}{L^2}
\overset{p}{\underset{k=1}{\Pi}}
d^2r'_{ki}\nonumber\\&& \eexp{[2\sigma\sum_{k=1}^{p_1}\sum_{k'=1}^p\sum_{i=2}^{n'}\ln\frac{L}{|\rr_{k1}-\rr_{k'i}|}+2\sigma\sum_{k,k'=1}^p\sum_{i<j}^{n'}\ln\frac{L}{|\rr_{ki}-\rr_{k'j}|}]+[\text{all}\,\, \rr_{ki}-\rr_{k'j}\rightarrow \rr'_{ki}-\rr'_{k'j}]-[\text{all}\,\, \rr_{ki}-\rr_{k'j}\rightarrow \rr_{ki}+\rr'_{k'j}]}\}\nonumber\\&& \eexp{2\sigma\ln\frac{L}{\ell_0}[\half p_1(p_1-1)+\half p(p-1)(n'-1)]\cdot 2}
\eeq
where $\{...\}$ is $L$ independent. Hence
\[ \frac{\ln Q_{1,-1}}{\ln L}=2+4\frac{n-p_1}{p}+2\sigma n+2\sigma[p_1(p_1-1)+p(p-1)\frac{n-p_1}{p}]\overset{n\rightarrow 0}{=}2-4\frac{p_1}{p}+2\sigma p_1(p_1-p) \]

Let us start with the case $\sigma > 2$.
Minimizing over $p$ yields $p=1/\sqrt{\sigma/2}$ and over $p_1$ is the same $p_1=1/\sqrt{\sigma/2}$. While there is RSB in the packet size $p$ there is no change in $Q_{1,-1}$, reflecting a $\sigma$ independent exponent,
\beq{65}
Q_{1,-1}\approx \left(\frac{L}{\ell_0}\right)^{-2}\ell_0^{-2}\qquad \sigma >0
\eeq
Indeed, optimizing now for $\sigma<2$ one finds $p=1$ and naively $p_1=(\half+1/\sigma)$.
However the optimum cannot be reached since one must have $p_1 \leq 1$,
hence
one must set $p=p_1=1$ leading to the same $-2$ exponent.

{\bf Remark}. One can check that the typical value of the ratio decays even faster than 
the average, i.e. 	
it can be argued to equals to 
$\log Z(\sqrt{2})- 2 \log Z(1)$ which leads to typical exponent 
$-2$ for $\sigma<1$, $4 (\sqrt{ \sigma}-1-\sigma/2)$ for $1<\sigma<2$
and $4 \sqrt{\sigma/2} (\sqrt{2}-2)$ for $\sigma>2$. 
\\

Consider next
\beq{66}
Q_{1,1}=\overline{ \frac{\int d^2r\, \eexp{2V(\rr)+2V(-\rr)}}{\int d^2r\,\eexp{2V(\rr)}\int d^2r\,\eexp{2V(-\rr)}}}
=\int d^2r
\overset{2n-2}{\underset{i=1}{\Pi}}
d^2r_i\,\overline{ \eexp{2V(\rr)+2V(-\rr)+2\sum_{i=1}^{2n-2} V(\rr_i)}}
\eeq
where in the denominator change variable so that $V(-\rr)\rightarrow V(\rr)$. Again, the pair $\rr, -\rr$ cannot be in the same packet since then $r\lesssim\ell_0$ and the c.m. of the packet cannot span the whole space. Consider then one packet $\rr_{k1},\,k=1,...,p_1$ with $p_1-1$ replicas and $\rr_{11}=\rr$, a second packet $\rr_{k2},\,k=1,...,p_2$ with $p_2-1$ replicas and $\rr_{12}=-\rr$ and then packets $\rr_{kj},\,k=1,...,p$ with $j=3,4,...,n'=\frac{2n-p_1-p_2}{p}+2$, so that the total number of replicas is $2n-2$.
\beq{67}
&&Q_{1,1}=\eexp{\sigma\ln\frac{L}{\ell_0}(2+2n-2)}L^{2+2(n'-2)}
\{\int\frac{d^2r}{L^2}
\overset{p_1}{\underset{k=2}{\Pi}}
\int d^2r_{k1}
\overset{p_2}{\underset{k=2}{\Pi}}
\int d^2r_{k2}
\overset{n'}{\underset{j=3}{\Pi}}
\frac{1}{L^2}
\overset{p}{\underset{k=1}{\Pi}}
\int d^2r_{kj}\nonumber\\&&
\eexp{2\sigma [\sum_{k=1}^{p_1}\sum_{k'=1}^{p_2}\ln\frac{L}{|\rr_{1k}-\rr_{2k'}|}
	+\sum_{k=1}^{p_1}\sum_{k'=1}^{p}\sum_{j=3}^{n'}\ln\frac{L}{|\rr_{1k}-\rr_{jk'}|}
	+\sum_{k=1}^{p_2}\sum_{k'=1}^{p}\sum_{j=3}^{n'}\ln\frac{L}{|\rr_{2k}-\rr_{jk'}|}
	+\sum_{k,k'=1}^{p}\sum_{3\leq i<j}^{n'}\ln\frac{L}{|\rr_{ik}-\rr_{jk'}|}]}\}
\nonumber\\&& \eexp{2\sigma\ln\frac{L}{\ell_0}[\half p_1(p_1-1)+\half p_2(p_2-1)+\half p(p-1)\frac{2n-p_1-p_2}{p}]}
\eeq
where $\{...\}$ is $L$ independent. Hence
\beq{68}
&&\frac{\ln Q_{1,1}}{\ln L}=\sigma[2n+p_1(p_1-1)+p_2(p_2-1) +(p-1)(2n-p_1-p_2)]
+2+2\frac{2n-p_1-p_2}{p}\nonumber\\&&
\overset{n\rightarrow 0}{=}\sigma[p_1^2+p_2^2-p(p_1+p_2)]+2-2\frac{p_1+p_2}{p}
\eeq
Minimizing with respect to $p,p_1,p_2$ yields $p=p_1=p_2=1$ for $\sigma<2$ while $p=p_1=p_2=1/\sqrt{\sigma/2}$ for $\sigma >2$. Thus also $Q_{1,1}$ is independent of $\sigma$,
\beq{69}
Q_{1,1}\approx \left(\frac{L}{\ell_0}\right)^{-2}\ell_0^{-2}\qquad \sigma >0
\eeq

Consider next the density overlap of two states displaced by $r_0\gg\ell_0$, supporting the form of $U_2$ in Eq. (17),
\beq{70}
Q(r_0)=\overline{ \frac{\int d^2r \eexp{2V(\rr)+2V(\rr-\rr_0)}}{[\int d^2r\eexp{2V(\rr)}]^2} } =\int d^2r
\overset{2n-2}{\underset{i=1}{\Pi}}
d^2r_i\,\overline{ \eexp{2V(\rr)+2V(\rr-\rr_0)+2\sum_{i=1}^{2n-2} V(\rr_i)}}
\eeq
The evaluation is similar to that of $Q_{1,1}$ with $-\rr$ replaced by $\rr-\rr_0$. Since $\rr_0\gg\ell_0$ the packets containing either $\rr$ or $\rr-\rr_0$ are distinct. Hence the derivation is the same as for $Q_{1,1}$ with the result $Q(\rr_0)\approx 1/L^2$. 

We note here that our sum rules in section IV are proven for large $\sigma$. In fact one can check the consistency of identifying $\epsilon$ as a chemical potential in Eq. (13) by substituting the result Eq. (17) in the first line of Eq. (13) and using $\overline{\rho(E,\rr)}$. This reproduces precisely Eq. (14) except for a factor $(1-\frac{1}{z})$, hence the derivation is valid for $z\gg 1$. 

We summarize here the various Coulomb matrix elements of Eq. (17) for the $E=0$ states and for all $\sigma$. These matrix elements serve as indicators for the cascade transitions also at $\sigma<2$.
	\beq{71}
	&U_1=U\ell_0^2P_{2,2}&\approx U \qquad\qquad\qquad\qquad\qquad \sigma>2\nonumber\\
	&\qquad &\approx U\left(\frac{L}{\ell_0}\right)^{-\frac{(2-\sigma)^2}{4\sigma}}\qquad\qquad \frac{2}{3}<\sigma<2 \nonumber\\
	&\qquad &\approx U\left(\frac{L}{\ell_0}\right)^{2\sigma-2}\qquad\qquad\,\,\,\,\,\,\,\, \sigma<\frac{2}{3}
	\nonumber\\&
	U_2=U\ell_0^2 Q_{1,\pm 1}&\approx \left(\frac{L}{\ell_0}\right)^{-2} 
	\qquad\qquad\qquad\,\,\,\,\,\,
	\mbox{all}\,\,  \sigma \nonumber\\&
	U_3=U\ell_0^2P_{2,0}&\approx \left(\frac{L}{\ell_0}\right)^{2-8\sqrt{\sigma/2}}
	\qquad\qquad\,\,\,\, \sigma>2\nonumber\\&
	\qquad &\approx \left(\frac{L}{\ell_0}\right)^{-2\sigma-2} \qquad\qquad\qquad \sigma<2
	\eeq
	Hence all matrix elements and the resulting cascade transitions are $\sigma$ independent at $\sigma>2$, in particular $U_3$ is vanishingly small. In contrast, at $\sigma<2$ the matrix element $U_1$ decreases while $U_3$ becomes comparable to $U_1,\,U_2$ so we expect that the form and strength of the cascade peaks change considerably. Thus the freezing transition, which is a minor effect in the DOS, becomes prominent when observing the cascade transitions.

\section{Disorder average $\overline{\rho^2_\epsilon}$}

We evaluate here the disorder average $\overline{\rho_\epsilon^2}$ which is used in Eq. \eqref{e14} to evaluate the density overlaps of all $E\neq 0$ states.
This is achieved by considering a 2-layer problem, following \cite{bh2} section IVB and \cite{bh} section IVB. The average DOS itself, summarized below Eq. (8), is a similar one-layer problem.
The layers $\uparrow,\,\downarrow$ are identical, for later use they have distinct fugacities $\epsilon_\uparrow.\,\epsilon_\downarrow$, so the interaction and the Gaussian terms, and the variational Hamiltonian are (eventually we replace $\epsilon_\uparrow,\,\epsilon_\downarrow \rightarrow \tilde\epsilon$ as defined below Eq. (6)),
\beq{81}
&&{\cal H}_{int}=\frac{1}{\pi\alpha}[\epsilon_\uparrow\cos\phi_\uparrow +\epsilon_\downarrow\cos\phi_\downarrow]\rightarrow \sum_{\{\nn_\uparrow,\nn_\downarrow\neq 0\}}\epsilon_\uparrow^{\,\sum_a n_{\uparrow a}^2}\epsilon_\downarrow^{\,\sum_a n_{\downarrow a}^2}
\eexp{i\nn_\uparrow\cdot{\bm\phi}_\uparrow+i\nn_\downarrow\cdot{\bm\phi}_\downarrow}
\nonumber\\&&
{\cal H}'=\int d^2r[\frac{1}{8\pi}\sum_{a,b}(K^{-1}\delta_{a,b}+2\sigma){\bm\nabla}\phi_a^+\cdot{\bm\nabla}\phi_b^+ +\frac{1}{8\pi K}{\bm\nabla}\phi_a^-\cdot{\bm\nabla}\phi_b^-]
\nonumber\\&& {\cal H}_0={\cal H}'+\half\int d^2r\sum_{a,b,\pm}(\sigma_c^\pm \delta_{ab}+\sigma_0^\pm)\phi_a^\pm\phi_b^\pm
\eeq
where $\phi^\pm=(\phi^\uparrow \pm\phi^\downarrow )/\sqrt{2}$ and the full Hamiltonian is ${\cal H}={\cal H}'-\sum_\rr {\cal H}_{int}$. Note that the disorder couples only to $\phi_a^+$ with strength $2\sigma$. The Greens' functions for ${\cal H}_0$ yield, defining $u_\pm, A_\pm$ as in \cite{bh}
\beq{82}
&&\int_q G_\pm =-4u_\pm\delta_{ab}-2A_\pm,\qquad\qquad u_\pm=\frac{K}{4}\ln(4\pi K\ell_0^2\sigma_c^\pm)\nonumber\\&&
A_+=-\sigma K^2\ln(4\pi K\ell_0^2\sigma_c^+) +K\frac{\sigma_0^+}{2\sigma_c^+} - K^2\sigma,\qquad\qquad A_-=K \frac{\sigma_0^-}{2\sigma_c^-}
\eeq
as in Eq. (58) of \cite{bh} and $1/\ell_0$ serves as a momentum cutoff. The interaction average becomes
\beq{83}
&&\langle {\cal H}_{int}\rangle_0=\sum_{\{\nn_\uparrow,\nn_\downarrow\neq 0\}}\epsilon_\uparrow^{\,\sum_a n_{\uparrow a}^2}\epsilon_\downarrow^{\,\sum_a n_{\downarrow a}^2}\langle \eexp{i\nn_\uparrow\cdot({\bm \phi}_+ +{\bm\phi}_-)/\sqrt{2}+i\nn_\downarrow\cdot({\bm\phi}_+ -{\bm\phi}_-)/\sqrt{2}}\rangle_0=\nonumber\\&&
\sum_{\{\nn_\uparrow,\nn_\downarrow\neq 0\}}\epsilon_\uparrow^{\,\sum_a n_{\uparrow a}^2}\epsilon_\downarrow^{\,\sum_a n_{\downarrow a}^2}
\eexp{u_+\sum_a(n_{\uparrow a}+n_{\downarrow a})^2 +\half A_+(\sum_a n_{\uparrow a}+n_{\downarrow a})^2+u_-\sum_a(n_{\uparrow a}-n_{\downarrow a})^2 +\half A_-(\sum_a n_{\uparrow a}-n_{\downarrow a})^2}=\nonumber\\&&
\sum_{\{n_{\alpha\beta}\}}\frac{m!}{n_{00}!n_{+0}!...n_{++}!}\langle 
\epsilon_\uparrow^{\,\sum_a n_{\uparrow a}^2}\epsilon_\downarrow^{\,\sum_a n_{\downarrow a}^2}\eexp{u_+\sum_a(n_{\uparrow a}+n_{\downarrow a})^2+u_-\sum_a(n_{\uparrow a}-n_{\downarrow a})^2+\omega_+\sum_a(n_{\uparrow a}+n_{\downarrow a})+\omega_-\sum_a(n_{\uparrow a}-n_{\downarrow a})}\rangle_\omega
\nonumber\\&&
\langle ...\rangle_\omega\equiv \int\int \eexp{-\frac{\omega_+^2}{2A_+}-\frac{\omega_-^2}{2A_-}}\frac{d\omega_+ d\omega_-}{2\pi\sqrt{A_+ A_-}}
\eeq
$n_{\alpha\beta}$ with $\alpha,\,\beta=\pm 1,0$ count the number of pairs that have $\alpha$ entry in the $\uparrow$ layer \underline{and} $\beta$ entry in the $\downarrow$ layer, $\sum_{\alpha\beta}n_{\alpha\beta}=m$.
All the replica sums in the exponent can be written in terms of $n_{\alpha\beta}$. The sum can be written as
\beq{84}
&& \sum_{\{n_{\alpha\beta}\}}\frac{m!}{n_{00}!n_{+0}!...n_{++}!}\langle 
(\epsilon_\uparrow\eexp{u_++u_-+\omega_++\omega_-})^{n_{+0}}
(\epsilon_\downarrow\eexp{u_++u_-+\omega_+-\omega_-})^{n_{0+}}
(\epsilon_\downarrow\eexp{u_++u_--\omega_++\omega_-})^{n_{0-}}
(\epsilon_\uparrow\eexp{u_++u_--\omega_+-\omega_-})^{n_{-0}}\nonumber\\&&
\times (\epsilon_{\uparrow}\epsilon_{\downarrow}\eexp{4 u_++2\omega_+})^{n_{++}}
(\epsilon_{\uparrow}\epsilon_{\downarrow}\eexp{4 u_-+2\omega_-})^{n_{+-}}
(\epsilon_{\uparrow}\epsilon_{\downarrow}\eexp{4 u_+-2\omega_+})^{n_{--}}
(\epsilon_{\uparrow}\epsilon_{\downarrow}\eexp{4 u_--2\omega_-})^{n_{-+}}\rangle_\omega
\eeq
To prove this, check the exponents of $\epsilon_{\uparrow},\epsilon_{\downarrow}$ and the coefficients of either of $u_\pm,\,\omega_\pm$ :
\beq{85}
&& \epsilon_\uparrow : n_{+0}+n_{-0}+n_{++}+n_{+-}+n_{--}+n_{-+}=\sum_a n_{\uparrow a}^2\nonumber\\&&
\epsilon_\downarrow: n_{0+}+n_{0-}+n_{++}+n_{+-}+n_{--}+n_{-+}=\sum_a n_{\downarrow a}^2\nonumber\\&&
u_+: n_{+0}+n_{0+}+n_{0-}+n_{-0}+4n_{++}+4n_{--}=\sum_a (n_{\uparrow a}+n_{\downarrow a})^2\nonumber\\&&
u_-: n_{+0}+n_{0+}+n_{0-}+n_{-0}+4n_{+-}+4n_{-+}=\sum_a (n_{\uparrow a}-n_{\downarrow a})^2\nonumber\\&&
\omega_+: n_{+0}+n_{0+}-n_{0-}-n_{-0}+2n_{++}-2n_{--}=\sum_a (n_{\uparrow a}+n_{\downarrow a})\nonumber\\&&
\omega_-: n_{+0}-n_{0+}+n_{0-}-n_{-0}+2n_{+-}-2n_{-+}=\sum_a (n_{\uparrow a}-n_{\downarrow a})
\eeq
This confirms \eqref{e83} so that the summation is a ninomial expansion
\beq{86}
&& Z=1+
\epsilon_\uparrow\eexp{u_++u_-+\omega_++\omega_-}+
\epsilon_\downarrow\eexp{u_++u_-+\omega_+-\omega_-}+
\epsilon_\downarrow\eexp{u_++u_--\omega_++\omega_-}+
\epsilon_\uparrow\eexp{u_++u_--\omega_+-\omega_-}+\nonumber\\&&
\epsilon_{\uparrow}\epsilon_{\downarrow}\eexp{4 u_++2\omega_+}+
\epsilon_{\uparrow}\epsilon_{\downarrow}\eexp{4 u_-+2\omega_-}+
\epsilon_{\uparrow}\epsilon_{\downarrow}\eexp{4 u_+-2\omega_+}+
\epsilon_{\uparrow}\epsilon_{\downarrow}\eexp{4 u_--2\omega_-}\nonumber\\
&&\frac{1}{m}\langle {\cal H}_{int}\rangle_0=\frac{1}{m} \langle Z^m\rangle_\omega
\overset{m\rightarrow 0}{=}
\langle \ln Z\rangle_\omega
\eeq
As in \cite{bh} we expect  the solution where $\sigma_c^-=0$ (i.e. it is irrelevant). Hence $u_-\rightarrow -\infty$ and 
\beq{87}
Z=1+\epsilon_{\uparrow}\epsilon_{\downarrow}\eexp{4 u_++2\omega_+}+
\epsilon_{\uparrow}\epsilon_{\downarrow}\eexp{4 u_+-2\omega_+}
\eeq
Denoting $\tilde \epsilon = \sqrt{\epsilon_{\uparrow}\epsilon_{\downarrow}}$, 
by a similar calculation as the one leading to (15)-(17) in \cite{bh} 
the self-consistent equation is now 
\be 
\sigma_c^+ \sim {\rm Prob}(2 u_+ + \omega_+ + \log \tilde \epsilon>0)
\sim \exp\left(  \frac{(K + 2 \frac{\log \tilde \epsilon}{\log \sigma_c^+})^2}{8 \sigma K^2} \log \sigma^+_c  \right)
\ee 
which means that $\sigma_c^+$ is dominated by the tail of the Gaussian distribution of $\omega_+$.
This leads to $\sigma_c^+ \sim \tilde\epsilon^{\,2/z}$   
with the previous $z=K(\sqrt{8\sigma}-1)$.
A stronger argument, valid more generally is to note (following \cite{bh}) that the solution is the same as in the one layer case with $K\rightarrow 2K$ and $ \epsilon_\downarrow\epsilon_\uparrow\rightarrow \tilde\epsilon^2$, 
hence 
$\sigma_c^+\sim(\tilde\epsilon^2)^{2/2z}=\tilde\epsilon^{\,2/z}$   
with the previous $z=K(\sqrt{8\sigma}-1)$. 

Our aim is to find the disorder average of $\rho_{\epsilon_\uparrow}
\rho_{\epsilon_\downarrow}$. Using the interaction average Eq. \eqref{e86} and the bosonic form of Eq. (7)
\beq{88}
&&\overline{\rho_\epsilon^2}=(\frac{1}{\pi\ell_0^2 })^{2}(\frac{\tilde\epsilon}{\epsilon})^2\frac{\partial^2}{\partial\epsilon_\uparrow\partial\epsilon_\downarrow}
\langle \ln [ 1+\epsilon_{\uparrow}\epsilon_{\downarrow}\eexp{4 u_++2\omega_+}+
\epsilon_{\uparrow}\epsilon_{\downarrow}\eexp{4 u_+-2\omega_+}]\rangle_\omega=
\nonumber\\&& (\frac{1}{\pi\ell_0^2 })^{2}(\frac{\tilde\epsilon}{\epsilon})^2
\langle \frac{\eexp{4 u_+ +2\omega_+}+\eexp{4 u_+ -2\omega_+}}
{[1+ \tilde\epsilon^2(\eexp{4 u_++2\omega_+}+\eexp{4 u_+-2\omega_+})]^2}\rangle_\omega
=(\frac{1}{\pi\ell_0^2 })^{2}\frac{\ell_0^2\sigma_c(2K,\tilde\epsilon^2)}{\epsilon^2}
\eeq
In the last equality we identify the mass parameter $\sigma_c$ of the DOS evaluation  as given in Eq. (15) of \cite{bh} and note that 
$\eexp{2u}$ is negligible in that expression. 
The result \eqref{e88} is reproduced in Eq. (14) which eventually leads to the various density overlaps as summarized in Eq. (17).

\end{widetext}


\begin{thebibliography}{10}
\bibitem{footnotetransverse}  Its longitudinal part can be gauged away as a random phase, immaterial in the following.
\end{thebibliography}
\end{document}